\documentstyle[preprint,eqsecnum,aps]{revtex}
\begin{document}
%%%%%%%%%%%%%%%%%%%%%%%%%%%%%%%%%%%%%%%%%%%%%%%%%%%%%%%%%%%%%%%%%%%%%%%%%%%%%%%%

%\def\ni{\noindent}
%\def\sw{Schwarzschild}

\title{Magnetic Field Decay in Neutron Stars. \\
       Analysis of General Relativistic Effects}

\author{Ulrich Geppert}
\address{Astrophysikalisches Institut Potsdam \\
         An der Sternwarte 16, 14482 \\
 	 Potsdam, GERMANY \\
	 e-mail: urme@aip.de}
\author{Dany Page}
\address{Instituto de Astronom\'{\i}a, UNAM \\
         Circuito de la Investigaci\'on Cient\'{\i}fica \\
         04510 Mexico D.F. MEXICO \\
         e-mail: page@astroscu.unam.mx}
\author{Thomas Zannias}
\address{Instituto de F\'{\i}sica y Matem\'aticas \\
         Universidad Michoacana SNH \\ 
         Morelia, Mich. 58040, MEXICO \\
         e-mail: zannias@ginette.ifm.umich.mx}

\maketitle

\begin{abstract}%%%%%%%%%%%%%%%%%%%%%%%%%%%%%%%%%%%%%%%%%%%%%%%%%%%%%%%%%%%
An analysis of the role of general relativistic effects on the decay of 
neutron star's magnetic field  is presented.
At first, a generalized induction equation on an arbitrary static background
geometry has been derived and, secondly, by a combination of analytical and
numerical techniques, a comparison of the time scales for the decay of an 
initial dipole magnetic field in flat and curved spacetime is discussed. 
For the case of  very  simple neutron star models, rotation not accounted for 
and in the absence of cooling effects, we find that the inclusion of general
relativistic effects result, on the average, in an enlargement of the decay 
time of the field
in comparison to the flat spacetime case. 
Via numerical techniques we show that, 
the enlargement factor depends upon the
dimensionless  compactness ratio ${\epsilon}={2GM \over c^{2}R}$, and 
for 
${\epsilon}$ in the range $(0.3~,~0.5)$, corresponding to 
compactness ratio of realistic neutron star models,
this factor is between  $1.2$ to $1.3$. 
The present analysis shows 
that general relativistic effects on the magnetic field decay ought to be examined 
more carefully than hitherto. 
A brief discussion of our findings on the impact of neutron stars physics is 
also presented.
\end{abstract}%%%%%%%%%%%%%%%%%%%%%%%%%%%%%%%%%%%%%%%%%%%%%%%%%%%%%%%%%%%%%%%%%%


%%%%%%%%%%%%%%%%%%%%%%%%%%%%%%%%%%%%%%%%%%%%%%%%%%%%%%%%%%%%%%%%%%%%%%%%%%%%%%%%
\section{Introduction}%%%%%%%%%%%%%%%%%%%%%%%%%%%%%%%%%%%%%%%%%%%%%%%%%%%%%%%%%%
%%%%%%%%%%%%%%%%%%%%%%%%%%%%%%%%%%%%%%%%%%%%%%%%%%%%%%%%%%%%%%%%%%%%%%%%%%%%%%%%

It is well known \cite{1} that a magnetic field in a plasma of finite conductivity 
is subject to diffusion and dissipation.
Diffusion leads to a spreading of inhomogeneities while dissipation is due to the 
Ohmic decay of the currents producing the field.
More concretely, a magnetic field ${\vec B(t,{\vec x})}$ in a plasma of uniform 
conductivity ${\sigma}$ evolves, in flat space-time, according to the following
diffusion equation \cite{1}:
\begin{equation}
{{\partial {\vec B(t,{\vec x})}\over {\partial t}}}=
{c^{2}\over 4{\pi}{\sigma}} \; \;  {\nabla}^{2}{\vec B(t,{\vec x})}.
\label{1}
\end{equation}
Accordingly, if $L$ is a typical length scale of the field structure, then it 
will decay-diffuse in a characteristic time  scale $\tau_{\rm Ohm}$ given by: 
$\tau_{\rm Ohm}=\frac{4 \pi \sigma L^2}{c^2}$.
Depending upon the prevailing conditions, the Ohmic decay time $\tau_{\rm Ohm}$ can
range from seconds, in the case of a copper sphere of radius of a few  centimeters \cite{1},
up to $\tau_{\rm Ohm}=10^{10}$ years or even much longer for astrophysical settings, 
as in the case of the sun \cite{1} or a neutron star \cite{2}.

The interactions of large scale cosmic magnetic fields with  plasmas is a problem of great 
importance in astrophysics and cosmology.
A particularly thorny issue nowadays concerns the origin and maintenance of cosmic magnetic 
fields.
Although large scale fields have been observed \cite{3}, a satisfactory explanation of 
their origin is still lacking.
Peebles \cite{4} considers the issue of the origin of the primordial magnetic field as one 
of the most important unsolved problems in cosmology. 
At the same time  the gigantic field of the pulsars begs for an explanation \cite{5}.
The general consensus of the astrophysical community \cite{6} is that such large scale 
fields have been generated via an episode of dynamo action 
\cite{7}, and then gradually suffer Ohmic decay due to the finite conductivity of the medium.
It appears therefore that an understanding of the factors influencing the decay of large 
scale fields, combined with relevant observations, may offers important clues towards a 
better understanding of the initial scale involved as well as clues regarding its origin.

In neutron stars the decay of the magnetic field is an issue of out most importance by 
itself \cite{8} and accordingly there has been an intense effort by astrophysicists to
understand the factors governing this decay.
As far as we are aware all theoretical modeling  of magnetic field decay in neutron stars 
utilized the familiar flat space time form of Maxwell's equations (an exception to this 
rule constitutes the recent work of ref. \cite{9}).
Although the employment of such framework is a fruitful one and provides us with valuable 
informations, it altogether neglects the background curvature of the spacetime which for 
the case of neutron stars is not any longer weak.
It would be worth to stress in that regard that curvature can modify considerably flat 
space-time solutions of Maxwell's equations. 
For instance, the reader may compare the solution describing a dipole magnetic field on a 
$Schwarzschild$ background \cite{10} to that of a flat space time. 
The presence of the logarithmic term in the former (see eqs.~\ref{23} further below) is a 
sole consequence of the non vanishing curvature. 
This example suggests that the role of the spacetime curvature on the decay process of 
magnetic fields ought to be examined more thoroughly than hitherto.
In that respect, we are aware only of the recent work of Sengupta, \cite{9}, where an 
investigation of  general relativistic  effects in the magnetic field decay of neutron 
stars have been attempted.
However this work is restricted to the study of magnetic fields confined only to the
outermost layers of a neutron star and furthermore it is assumed  that those outermost 
layers (and thus also  the magnetic field  ${\vec B}$), are embedded on a $Schwarzschild$ 
background geometry.
Thus strictly the framework of ref.\cite{9}, deals exclusively with magnetic decay
on a $Schwarzschild$ background.
In addition to those approximations and according to the sentence following Eq. 15 
of Sengupta's second work, the  author fails to include general relativistic effects on 
the outer boundary condition for matching the inner field with the outer vacuum dipolar
field across the surface of the star.
In contrast, in the present work, a broad framework dealing with general relativistic
effects on the magnetic field decay on an arbitrary static geometry and  with proper 
allowance of the correct general relativistic inner and outer boundary conditions is 
presented.
Moreover, and in contrast to the approach of ref. \cite{9}, we formulate the entire
problem avoiding the introduction of a vector potential and the associated ambiguities.
Our analysis shows that general relativistic effects \cite{11}
can influence  the field decay,
but the precise manner that this influence manifests itself depends upon the class of
observers called in to describe the field decay. 
For the magnetic field of a non rotating neutron star it is natural to describe the 
field decay relative to the class observers that find themselves at rest relative
to the star, ie the class of Killing observers. 
Relative to such observers, we find that relativistic effects are influencing  
the field decay via two major modes:
the gravitational red shift as well as the intrinsic curved geometry of the 
spatial sections constituting the rest space of the Killing observers.
Subsequent numerical analysis shows that the red shift factor 
is the dominant one 
in slowing down the field decay. 
Overall we find that the inclusion of relativistic effects 
make the decay time of the
field larger than, but of the same order of magnitude,
 as in flat space-time. Nevertheless the
preliminary study of the present paper utilizing a simple non rotating neutron star models
suggests that general relativistic effects should be given further considerations.
We explicitly illustrate the impact of relativistic effects upon the magnetic field decay,
by examining  the evolution of a magnetic field permeating a constant density neutron star,
first in their presence and secondly without them.
Although for both treatments we have  obtained exponential decays,  the decay time in the 
presence of relativistic effects, on the average, is enlarged by a factor
that
depends crucially upon the value of the compactness 
ratio ${\epsilon}={2GM\over c^{2}R}$.
Specifically for values of ${\epsilon}$ in the domain $(0.3~,~0.5)$, 
characterizing
realistic neutron star models, we find that the decay time
is $1.2$ to $1.3$ larger than the corresponding flat decay time,
while for higher values of ${\epsilon}$, it can be
 larger. We may add parenthetically that the term 
"average" increase in the decay time, is explained in details 
in section (IV) of the
paper. 
 
The present paper is organized as follows: In the following section, starting from 
Maxwell's equations on a static spacetime we first derive the relevant induction equation 
taking  into account the curved nature of the background spacetime geometry.
It should be stressed however that the employment of a static  geometry does not leave
room for incorporating gravitomagnetic (Lense-Thirring)
effects in the induction equation, as the latter
would manifest themselves relative to non static backgrounds, but we do hope to
present such analysis  in a future work.
In section III, we specialize the induction equation to a simple neutron star model and 
a detailed analysis of the content of the induction equation is presented.
In the same section the sensitive issue of the boundary conditions accompanying the 
induction equation is also addressed. 
In the section IV, we  discuss numerical solutions of the curved spacetime induction equation 
and an assessment of the relativistic factors influencing the field decay is discussed.
Furthermore in the same section, a comparison of the field decay in curved and 
flat spacetime is also presented.
In the concluding section,  a brief discussion of the physical implications of our results 
to neutron stars physics is presented and possible extension of the present work is outlined.
Finally we have included an Appendix where a few  intermediate calculations leading to the 
main equations of section II are presented.

%%%%%%%%%%%%%%%%%%%%%%%%%%%%%%%%%%%%%%%%%%%%%%%%%%%%%%%%%%%%%%%%%%%%%%%%%%%%%%%%
\section{Induction equation on a static background geometry}%%%%%%%%%%%%%%%%%%%%
%%%%%%%%%%%%%%%%%%%%%%%%%%%%%%%%%%%%%%%%%%%%%%%%%%%%%%%%%%%%%%%%%%%%%%%%%%%%%%%%

Maxwell's equations, in covariant form, are as follows \cite{12,13}:
\begin{mathletters}
\label{2}
\begin{equation}
{\nabla}^{\alpha}F_{\alpha\beta}=-{4{\pi}\over c}J_{\beta}
\label{2a}
\end{equation}
\begin{equation}
{\nabla}_{[{\alpha}}F_{{\beta\gamma}]}=0
\label{2b}
\end{equation}
\end{mathletters}
where $F_{\alpha\beta}=-F_{\beta\alpha}$, $J_{\alpha}$ and ${\nabla}$ are the coordinate
components of the Maxwell tensor, the  conserved four current and the derivative operator 
respectively.
Given a solution $F_{\alpha\beta}$ of the above eqs., an observer with four velocity  
$U^{\alpha}$, $U^{\alpha}U_{\alpha}=-1$, measures electric and magnetic fields 
$(E,B)$ with  corresponding coordinate components given respectively by:
\begin{equation}
E_{\alpha}=F_{\alpha\beta}U^{\beta}~~,~~B_{\alpha}=-
{1\over 2}{{\epsilon}_{\alpha\beta}}^{\gamma\delta}
F_{\gamma\delta}U^{\beta}
\label{3}
\end{equation}
where ${\epsilon}_{\alpha\beta\gamma\delta}$ stands for the four-dimensional Levi-Civita
tensor density \cite{14}.
We shall be concerned in the present paper with particular solutions of \ref{2} where the 
current $J$ is described by the following relativistic extension of Ohm's law,
as it was first formulated by Weyl \cite{15}:
\begin{equation}
J^{\alpha}={\sigma}g^{\alpha\beta}F_{\beta\gamma}V^{\gamma}
\label{4}
\end{equation}
where in above equation, $(V^{\alpha},{\sigma})$ stand for the four velocity of a conducting neutral 
plasma and  its scalar electrical conductivity \cite{16} respectively.
Although eqs.~\ref{2} to \ref{4} are valid for any kind of background geometries and plasmas 
characterized by arbitrary four velocity and conductivity, hereafter we shall 
restrict our consideration to background geometries that are globally static. 
Staticity in turn allows us to select coordinates so that the spacetime geometry can be 
written in the form (see for instance discussion in ref. \cite{12,13}):
\begin{equation}
ds^{2}=-e^{2{\Phi}}(dx^{o})^{2}+{\gamma}_{ij}dx^{i}dx^{j}
\label{5}
\end{equation}
where $x^{o}=ct$, ${\gamma}_{ij}$ are functions of the the spatial coordinates 
$x^{i}$, $(i=1,2,3)$, and by ${\xi}$ denote the hypersurface orthogonal timelike Killing
vector field obeying:
${\xi}_{\alpha}{\xi}^{\alpha}=-e^{2{\Phi}}$.
For the above form of the line element, Maxwell's equations \ref{2} and the current 
conservation law  ${\nabla}_{\alpha}J^{\alpha}=0$ can be re-written in an equivalent 
form involving only the components $(E^{i},B^{i})$ of the electric and magnetic fields 
respectively, as well as the charge density $c{\rho}=-U_{\mu}J^{\mu}$ and spatial current
density $J^{i}$ as measured by the Killing observers \cite{17,18}. 
More precisely if by $U^{\mu}$ we denote their four velocity then eqs.~\ref{2} yield 
the following equivalent set (see Appendix for details, or ref.~\cite{17,18}):
\begin{mathletters}
\label{6}
\begin{equation}
D_{i}E^{i}=4{\pi}{\rho}~~~~~~~~~~,~~~~~~~~~~D_{i}B^{i}=0
\label{6a}
\end{equation}
\begin{equation}
{\epsilon}^{ijk}D_{j}(ZB_{k})=
{4{\pi}\over c}ZJ^{i}+{\partial E^{i}\over {\partial x^{o}}}
\label{6b}
\end{equation}
\begin{equation}
{\epsilon}^{ijk}D_{j}(ZE_{k})=
-{\partial B^{i}\over {\partial x^{o}}}
\label{6c}
\end{equation}
\begin{equation}
U^{\mu}{\partial (c{\rho})\over {\partial x^{\mu}}}+D_{i}J^{i}+J^{i}D_{i}
logZ=0
\label{6d}
\end{equation}
\end{mathletters}
where in above $D$ stands for the covariant derivative operator associated with 
${\gamma}$, ${\epsilon}^{ijk}$ represent the (coordinate) components of the three dimensional 
totally antisymmetric Levi-Civita  tensor density defined on the $x^{o}=const$ slices and 
$Z=(-{\xi}^{\alpha}{\xi}_{\alpha})^{1\over 2}=e^{\Phi}$ is the red shift factor which in 
the language of in the ${3+1}$ approach to spacetime or (and) electrodynamics, is also 
refered as the lapse function \cite{18}. 

With Maxwell's eqs. in the above form we can derive an induction equation by repeating
the same steps leading to the derivation of its flat counterpart (see for example 
discussion in \cite{1}).
For a plasma at rest relative to the Killing observers, combined with Ohm's law and the 
MHD approximation (i.e. neglecting the displacement current \cite{19} from the right hand 
side of \ref{6b}), one obtains from \ref{6}-abc the following form of the generalized 
induction equation:
\begin{equation}
{\partial B^{i}\over {\partial {x^{o}}}} +{\epsilon}^{ijk}D_{j}
\left[{c\over 4{\pi}{\sigma}}
{{\epsilon}_{k}}^{lm}D_{l}(ZB_{m})\right]=0
\label{7}
\end{equation}

This last equation describes the time evolution of a magnetic field configuration that 
finds itself in a conducting medium. 
In principle one could write down the explicit form of the dynamical evolution equation  
once a choice of  background geometry has been made. 
However before we do so, we would like to make a further specialization of the eqs.~\ref{6} 
and \ref{7}, so that their interrelationship to the familiar flat space three plus one
formalism of Maxwell's eqs. is  more transparent. 
Here, following the spirit of \cite{18} and particularly \cite{20}, we shall sacrifice the 
manifest three-covariance of eqs.~\ref{6} and \ref{7} with respect to arbitrary coordinate 
transformations of the $t=const$ sections, for the benefits of  practical usefulness.
As was pointed out in ref.\cite{18,20}, if one defines suitably the components of $(E,B)$ 
and under some weak constraints upon the spacetime geometry, then Maxwell's equations can 
be recast in a more "user friendly" form.
This new form employs concepts familiar from the language of the three dimensional vector 
analysis expressed in orthogonal curvilinear coordinates and such approach to curved 
spacetime electrodynamics is particularly useful  for astrophysical purposes. 
Having in mind further astrophysical applications of our results we shall recast 
eqs.~\ref{6}-abc in such a form.
Such form requires  that the geometry of the spacetime permits the introduction of 
coordinates so that the spatial three element $ds^{2}_{(3)}$ of \ref{5} could be recast in 
the following form:
\begin{equation}
ds^{2}_{(3)}=h_{1}^{2}(dx^{1})^{2}+h_{2}^{2}(dx^{2})^{2}+
h_{3}^{2}(dx^{3})^{2}
\label{8}
\end{equation}
where the scale factors $h_{i}=h_{i}(x^{1},x^{2},x^{3})$ are for the moment arbitrary 
functions of  $(x^{1}, x^{2}, x^{3})$. 
In the Appendix, (see also \cite {18,20}) we show that for such geometries 
eqs.~\ref{6} can be written in the 
following form:
\begin{mathletters}
\label{9}
\begin{equation}
{\nabla}{\cdot}{\vec E}=4{\pi}{\rho}~,~~~
{\nabla}{\cdot}{\vec B}=0
\label{9a}
\end{equation}
\begin{equation}
{{\vec \nabla}}{\times}(Z{\vec B})=
{4{\pi}\over c}Z{\vec J}+{1\over c}{\partial {\vec E}\over {\partial t}}
\label{9b}
\end{equation}
\begin{equation}
{{\vec \nabla}}{\times}(Z{\vec E})=
-{1\over c}{\partial {\vec B}\over {\partial t}}
\label{9c}
\end{equation}
\begin{equation}
{\nabla}{\cdot}{\vec J}+{\vec J}{\cdot}{\nabla}(logZ)=0
\label{9d}
\end{equation}
\end{mathletters}
where we have written the current conservation law for an electrically neutral plasma 
and in above equations the symbols $({\nabla}{\cdot}~,~{\vec \nabla}{\times}~,~ {\nabla})$
stand for the divergence, curl and gradient operators respectively, expressed entirely 
in terms of the scale factors $h_{i}$ (see Appendix for their explicit representation).
We also remind the reader that all vector components in equations \ref{9} are physical 
frame components taken with respect to the field of orthonormal frames 
$e_{i}={1\over h_{i}}{\partial \over {\partial x^{i}}},~(i=1,2,3)$,
naturally singled out by the line element \ref{8}.
Using now eqs.~\ref{9}, or directly from eq.~\ref{7}, upon eliminating the coordinate 
components of ${\vec B}$ in favor of its frame components, the induction equation \ref{7}
takes the following form:
\begin{equation}
{1\over c}{\partial {\vec B}\over {\partial t}}+ {\vec \nabla}{\times} 
\left[{c\over 4{\pi}{\sigma}}{\vec \nabla}{\times}(Z{\vec B})\right]=0
\label{10}
\end{equation}

Equations~\ref{9} and \ref{10} are the main equations of this section.
In the special case of a $Schwarzschild$ background, naturally they are reduced to those 
of ref.\cite{20}, and in the case of a plasma of uniform conductivity the generalized
induction eq.~\ref{10} reduces to eq.~\ref{1} in the limit of flat space. 

%%%%%%%%%%%%%%%%%%%%%%%%%%%%%%%%%%%%%%%%%%%%%%%%%%%%%%%%%%%%%%%%%%%%%%%%%%%%%%%%
\section{Magnetic field decay interior to neutron stars}%%%%%%%%%%%%%%%%%%%%%%%%
%%%%%%%%%%%%%%%%%%%%%%%%%%%%%%%%%%%%%%%%%%%%%%%%%%%%%%%%%%%%%%%%%%%%%%%%%%%%%%%%

In the neutron star's interiors the MHD approximation is well justified \cite{19} and
we shall explore the content of the relativistic induction equation \ref{10},
by applying it to study the evolution of magnetic fields associated with neutron stars.
Since the main purpose of the present work is to investigate the impact of the 
spacetime curvature upon  the magnetic field decay, as a first preliminary step
we shall adopt a rather simplified  neutron star model. 
The chosen model primarily avoids technicalities that may obscure the issue at hand but
at the same time it shows clearly the potential impact of the curvature on the magnetic
field decay.
Accordingly, and to avoid laborious numerical computations, we shall ignore the rotation 
of the neutron star and thus shall adopt as the background geometry a non-singular, 
static and spherically symmetric one.
Hence, the scale factors of eq.~\ref{8} will be taken as:
\begin{equation}
h_{r}^{2}={\left(1-{2Gm(r)\over rc^{2}}\right)^{-1}}
         ={\left(1-{2M(r)\over r}\right)^{-1}}~~,~
~h_{{\theta}}^{2}=r^{2}~~,~~
h_{\phi}^{2}=r^{2} \sin^{2}{\theta}
\label{11}
\end{equation} 
while for the moment the lapse or red shift factor $Z=Z(r)=e^{{\Phi}(r)}$ and 
the "mass function" $m=m(r)$ are arbitrary functions of the radial coordinate. 

We shall begin our analysis of the magnetic field decay by assuming 
that at some initial 
time $t_{o}$ an axially symmetric distribution of a magnetic field 
${\vec B}(t_{o}, r, {\theta})$ permeates the entire star. 
We are not concerned here upon the mechanism that brought such a field 
into existence 
but rather we are interested in its evolution. 
Its evolution is considerably affected by the electrical conductivity 
${\sigma}$, but
as a part of the adopted simplified picture and in order to emphasize the 
effects of space-time
curvature we shall take ${\sigma}$ to be spherically 
symmetric and shall ignore any cooling effects that may influence its 
temporal evolution.
For an axially symmetric field ${\vec B}$, it is convenient to decompose 
it into the so 
called poloidal ${\vec B}_{(p)}$ and toroidal part ${\vec B}_{(t)}$. 
In terms of the orthonormal basis vectors $(e_{r},~e_{\theta},~e_{\phi})$ 
those parts are 
defined respectively by:
${\vec B}_{(p)}=B^{r} \vec{e}_{r}+B^{\theta} \vec{e}_{\theta}$ and 
${\vec B}_{(t)}=B^{\phi} \vec{e}_{\phi}$
with $ (B^{r},B^{\theta}, B^{\phi}) $ arbitrary functions of $(t,r,{\theta})$ 
respectively. 
One can  then easily conclude from the induction eq.~\ref{10} that, 
as long as the scalar 
conductivity is spherically symmetric, the toroidal and poloidal 
parts of ${\vec B}$, 
evolve independently of each other \cite{21}. 
Such decoupling is rather convenient since it implies that if the 
initial distribution 
of the magnetic field is purely poloidal then it will not develop a 
toroidal component in 
the course of its evolution and vice versa. 
For simplicity, in the present paper we shall examine the effects of the spacetime curvature 
only on the  evolution of a purely poloidal field 
${\vec B}_{(p)}=B^{r} \vec{e}_{r}+B^{\theta} \vec{e}_{\theta}$.
For such field ${\vec B}$, it follows from \ref{9b} that the current ${\vec J}$ is along the 
$\vec{e}_{\phi}$ direction, and thus the current conservation eq.~\ref{9d} is identically 
satisfied. 
Furthermore via Ohm's law, and use of \ref{9b} (with the displacement current ignored), it 
follows that the electric field ${\vec E}$ is a purely toroidal and axisymmetric field and 
Gauss law ${\nabla}{\cdot}{\vec E}=0$ is satisfied as well.
Consequently, from the system of eqs.~\ref{9}, we are left to satisfy the constraint 
${\nabla}{\cdot}{\vec B}=0$, solutions of which will be evolved by  the induction eq.~\ref{10}. 

Taking into account the poloidal and axisymmetric nature of $\vec{B}$ as well as the formula
of the div operator ${\nabla}{\cdot}$, listed in the Appendix,  in view of the scale factors 
of \ref{11}, one easily finds  that  ${\nabla}{\cdot}B=0$ implies:
\begin{equation}
\left(1-{2M\over r}\right)^{1\over 2}{1\over r}{\partial (r^{2}B^{r})\over {\partial r}}
+{1\over \sin{\theta}}{\partial (B^{\theta} \sin{\theta})\over {\partial {\theta}}} =0
\label{12}
\end{equation}
We shall look for separable solutions of the above equations in the form:
\begin{equation}
B^{r}=F(t,r){\Theta_{1}}({\theta}),~~~B^{\theta}=G(t,r){\Theta_{2}}({\theta}) 
\end{equation}
with the functions $F,~G,~{\Theta_{1}},~{\Theta_{2}}$ to be determined.
Substituting the above representations of $(B^{r},~B^{\theta})$ in \ref{12} and separating 
variables one gets the following equivalent system:
\begin{mathletters}
\label{13}
\begin{equation}
\left(1-{2M\over r}\right)^{1\over 2}{1\over r}{\partial (r^{2}F)\over {\partial r}}-
{\lambda}G=0
\label{13a}
\end{equation}
\begin{equation}
{1\over \sin{\theta}}{\partial (\sin{\theta}\; {\Theta}_{2})\over 
{\partial {\theta}}}
+{\lambda}{\Theta}_{1}=0
\label{13b}
\end{equation}
\end{mathletters}
where ${\lambda}$ stands for a separation constant.
The second equation can be solved in terms of the Legendre polynomials by taking
${\lambda}=l(l+1),~~ l=0,1,2...$, and:
\begin{mathletters}
\label{14}
\begin{equation}
{\Theta}_{2}= \sin{\theta}{dP_{l}(y)\over dy},~~~{\Theta}_{1}=-P_{l}(y),~~~
y= \cos{\theta}
\label{14a}
\end{equation}
On the other hand, for such ${\lambda}$, eq.~\ref{13a} is satisfied provided, for 
$l {\neq} 0$, one chooses $G(r,t)$ in the following form;
\begin{equation}
G(t,r)={1\over l(l+1)}\left(1-{2M\over r}\right)^{1\over 2}{1\over r}
{\partial (r^{2}F)\over {\partial r}}
\label{14b}
\end{equation}
\end{mathletters}
We shall disregard the $l=0$ mode since, as it is clear form above, it corresponds to a 
monopole field $B$. 
With the exclusion of monopole fields, the components of an arbitrary axisymmetric poloidal
field can be written as a superposition of "l-poles" in the form:
\begin{mathletters}
\label{15}
\begin{equation}
B^{r}(t,r,{\theta})=-{\sum}^{\infty}_{l=1} \; F_{l}(t,r) \, P_{l}(y)
\label{15a}
\end{equation}
\begin{equation}
B^{\theta}(t,r,{\theta})={\sum}^{\infty}_{l=1} \; {1\over l(l+1)}
\left(1-{2M\over r}\right)^{1\over 2} \, {1\over r}
{\partial (r^{2}F)\over {\partial r}}\, \sin{\theta}\, {dP_{l}(y)\over dy}
\label{15b}
\end{equation}
\end{mathletters}
To simplify algebra, and on physical grounds, we shall restrict our considerations to the 
detailed analysis of only the $l=1$ mode. 
Such mode corresponds to a dipole field and such configuration is expected to be present
and dominant within neutron stars.
For $l=1$, eqs.~\ref{15} yield:
\begin{equation}
B^{r}(t,r,{\theta})=-F(t,r)\, \cos{\theta},~~B^{\theta}(t,r,{\theta})=
{1\over 2r}
\left(1-{2M\over r}\right)^{1\over 2} \,
{\partial (r^{2}F)\over {\partial r}} \, \sin{\theta}
\label{16}
\end{equation}
where for notational simplicity we  write here after $F$ instead of $F_{1}$.
On the other hand, for any poloidal axisymmetric field with components $(B^{r},B^{\theta})$,
the induction equation \ref{10} on the background geometry of \ref{8}, yields the following
two non-trivial evolution equations:
\begin{mathletters}
\label{17}
\begin{equation}
{\partial B^{r}\over {\partial x^{o}}}+{1\over h_{\theta}h_{\phi}}
{\partial\over {\partial {\theta}}}\left[{cA\over 4{\pi}{\sigma}}\right]
=0
\label{17a}
\end{equation}
\begin{equation}
{\partial B^{{\theta}}\over {\partial x^{o}}}-{1\over h_{r}h_{\phi}}
{\partial\over {\partial {r}}}\left[{cA\over 4{\pi}{\sigma}}\right]
=0
\label{17b}  
\end{equation}
where:
\begin{equation}
A= {h_{\phi}\over h_{r}h_{\theta}}
\left[ {\partial \over {\partial r}} 
      \left( h_{\theta}B^{\theta}Z \right)-
               {\partial \over {\partial {\theta}}}
      \left(h_{r}B^{r}Z \right)
\right]
\label{17c}
\end{equation}
\end{mathletters}
When one now inserts in eq.~\ref{17a} the explicit forms of the components of 
$(B^{r},B^{\theta})$ corresponding to a dipole field in the form shown in eq.~\ref{16}, 
as well as the  scale factors of eq.~\ref{11}, then  gets:
\begin{equation}
{4{\pi}{\sigma}\over c}{\partial { F}\over {\partial x^{o}}}=
\left(1-{2M\over r}\right)^{1\over 2}{1\over r^{2}}
{\partial \over {\partial r}}
\left[Z\left(1-{2M\over r}\right)^{1\over 2}{\partial {(r^{2}F)}\over {\partial r}}\right]-
{2ZF\over r^{2}}
\label{18}
\end{equation}
In arriving at the above equation we have taken explicitly into account the spherically 
symmetric nature of the scalar conductivity ${\sigma}$. 
We may point out that for non spherical  ${\sigma}$, the right hand side of \ref{18} 
contains gradients of ${\sigma}$ along the meridian directions but for our simple 
neutron star model a spherical conductivity is rather adequate.
On the other hand identical manipulations of (3.8b) leads to:
\begin{mathletters}
\label{19}
\begin{equation}
{\partial \over {\partial r}}
\left[r^{2}{\partial F \over {\partial x^{o}}}-
      {c\over {4{\pi}{\sigma}}}\left(1-{2M\over r}\right)^{1\over 2}
      {\partial \over {\partial r}}
           \left[Z\left(1-{2M\over r}\right)^{1\over 2}
                 {\partial {(r^{2}F)}\over {\partial r}}\right]+
      {2{c\over {4{\pi}{\sigma}}}ZF}\right]=0
\label{19a}
\end{equation}
from which we infer that 
\begin{equation}
r^{2} { {\partial{F}} \over {\partial x^{o}} } -
{c \over {4{\pi}{\sigma}}} \left(1-{2M\over r}\right)^{1\over 2}
{\partial \over {\partial r}}
\left[Z\left(1-{2M\over r}\right)^{1\over 2}
 {\partial {(r^{2}F)} \over {\partial r}}\right] + 
2{c\over {4{\pi}{\sigma}}}ZF
=g({\theta},{\phi},t)
\label{19b}
\end{equation}
\end{mathletters}
where $g({\theta},{\phi},t)$ an integration "constant".
A comparison then between (3.9) and (3.10b) shows that necessary $g=0$.
If we further define: ${\hat F}=r^{2}F$, then one finds either from \ref{18} or \ref{19b}
that ${\hat F}$ satisfies:
\begin{equation}
{4{\pi}{\sigma}\over c}{\partial {\hat F}\over {\partial x^{o}}}=
\left(1-{2M\over r}\right)^{1\over 2}{\partial \over {\partial r}}
\left[Z\left(1-{2M\over r}\right)^{1\over 2} {\partial {\hat F}\over {\partial r}}\right]-
2{Z\over r^{2}}{\hat F}
\label{20}
\end{equation}

The above equation essentially describes the evolution of the the dipole field components 
\cite{22} as they are measured by the Killing observers. 
Linearity of Maxwell's and induction equation implies that \ref{20} specifies a unique 
solution up to an arbitrary rescaling. 
This rescaling freedom will be fixed  later on by a suitable matching of the interior 
dipole field to a corresponding asymptotically vanishing exterior dipole one. 

Taking $Z=1,~~M=0$ in eq.~\ref{20} one recovers the standard equation describing the 
evolution of a dipole axisymmetric poloidal field in flat space \cite{23} namely:
\begin{equation}
{4{\pi}{\sigma}\over c^{2}}{\partial {S}\over {\partial t}}=
{{\partial}^{2} S\over {\partial r^{2}}}-{2S\over r^{2}}
\label{21_0}
\end{equation}
where in order to avoid confusion we have denoted the analogue of ${\hat F(t,r)}$ for 
the flat space time case by $S(t,r)$.
The latter function  often in the astrophysics literature is referred to as the Stoke's 
function \cite{24}.
In order to get some insights into the significance of the various terms appearing in (3.11),
the 
general relativistic counterpart of \ref{21_0}, we shall rewrite the former in an equivalent
form so that a clean comparison between the two could be afforded.
Eliminating the (areal) radial coordinate $r$ in favor of the physical proper radius  
${l(r)}$ of the $r=constant$ spheres, via  ${dl={dr(1-{2M\over r})^{-{1\over 2}}}}$,
Eq.~\ref{20} takes then the following form:
\begin{mathletters}
\label{21}
\begin{equation}
{4{\pi}{\sigma}\over c^{2}}{\partial {\hat F}\over {\partial t}}=
{\partial \over {\partial l}}
(Z
{\partial {\hat F}\over {\partial l}})-
{2Z\over r(l)^{2}}{\hat F}
\label{21a}
\end{equation}

A comparison then to eq.~\ref{21_0} shows that relative to the Killing observers,
general relativistic effects can influence B-decay in three ways.
Namely via the presence of the red shift factor $Z$, its gradient and as well as via the
intrinsically curved nature of the rest space of the Killing observers, ie the $t=const$ 
hyperfaces. 
The latter manifest itself in \ref{21a} via the term $r(l)$,
a term which in 
general satisfies $r(l){\neq}l$, implying that 
that the rest spaces of the Killing observers are intrinsically
curved. 
From the above mentioned  three factors the spatial gradient of $Z$ makes negligible 
contribution to the field decay and this has been verified numerically.
Neglecting this gradient then equation \ref{21a} takes the following form:
\begin{equation}
{4{\pi}{\sigma}\over c^{2}Z(l)}{\partial {\hat F}\over {\partial t}}=
{{\partial}^{2} {\hat F}\over {\partial l^{2}}}
-{2{\hat F}\over r(l)^{2}}
\label{21b}
\end{equation}
\end{mathletters}
In this form a clean comparison to equation (3.12) can be afforded. 
The right hand sides of the two equations involve 
physical spatial gradients and the differ only  
 by terms of the order $O({2Gm\over c^{2}R})$.
On the other hand their left hand sides as they stand 
cannot be compared. If however one  reasonably
replaces $Z(l)$ by some averaged value $<Z>$,
then the left hand side of (3.13) involves also
physical temporal gradients. In that event
one gets a first flavor of the magnitude of the general relativistic effects.
They  modify the corresponding
flat spacetime results by terms of order unity.
Of course such conclusions has to be also documented 
at the solution level as well, and as we shall further ahead this indeed is the case.

Having thus identified the manner by which relativistic gravity effects the magnetic field 
decay, our assignment is now to access the relative importance of each of the above two
factors.
In the following section we shall do so by resorting to  numerical computations. 
However,  before we pass to that issue let us first record the suitable boundary conditions 
to be imposed upon the corresponding  $S(t,r)~,~{\hat F(t,r)}$ in order to describe
sensible physics. 
The required conditions for both, ie the flat and the general relativistic case, are 
drawn by demanding that the interior ${\vec B}$ ought to be a non singular field at
all times and at all spatial points, and in addition it ought to join smoothly across the 
surface of the star to an exterior asymptotically vanishing dipole field.
For the flat space case, taking  $M=0$ and $F(t,r)=-{{2S}\over r^{2}}$ in eq.~\ref{16} 
one gets: 
${\vec B}= {{2S}\over r^{2}} \cos{\theta}   \; \vec{e}_{r}-{1\over r}
{\partial S\over {\partial r}} \sin{\theta} \; \vec{e}_{\theta}$ 
from which we infer that the magnitude ${\vec B}^{2}$ of the interior magnetic field 
is given by: 
$|{\vec B}|^{2}={4S^{2}\over r^{4}} \cos^{2}{\theta}+{1\over r^{2}}
({\partial S\over {\partial r}})^{2} \sin^{2}{\theta}$. 
Accordingly a regular field at the star's center requires  $lim{S(t,r)\over r^{2}}$ to be 
finite as the center of the star is approached.
On the other hand an asymptotically vanishing dipole magnetic field in flat space due to 
a magnetic moment ${\mu}$, is described by \cite{1}: 
${\vec B}={2{\mu} \cos{\theta}\over r^{3}}\; \vec{e}_{r}+
{{\mu} \sin{\theta}\over r^{3}} \; \vec{e}_{\theta}$.
It follows then from the above expressions that a $C^{0}$ matching of the interior magnetic 
field to slow varying exterior dipole field \cite{25}, requires that across the star's 
surface, ie at radius $R$, $S(t,r)$ should satisfy:
\begin{equation}
\left. R \; {\partial S\over {\partial r}}\right|_{R}=-S
\label{22a}
\end{equation}
The relatively simple nature of (3.12) as well the simple form of the boundary-regularity 
conditions outlined above, permit us to construct exact closed form solutions.
In fact, it is not difficult to verify that for the case of a star of a uniform 
conductivity ${\sigma}$, a sequence of  exact solutions of eq.~(3.12) obeying the above 
described conditions is given by \cite{26}: 
\begin{equation}
S_{n}(t,x)=\left[{ \sin(n{\pi}x)\over {n^{2}{\pi}^{2}x}}-
{\cos(n{\pi}x)\over n{\pi}}\right] \; e^{-{t\over {\tau}_{n}}}=
{\psi}_{n}(x)\; e^{-{t\over {\tau}_{n}}}
\label{22b}
\end{equation} 
where 
${\tau}_{n}={4{\sigma}R^{2}\over {\pi}c^{2}n^{2}} = 
\frac{1}{n^2} \, \frac{\tau_{\rm Ohm}}{\pi^2}$, 
$x={r\over R}$ and  $n$ takes the values $(1,2,3....)$.

The above sequence of exact solutions offers a clear picture regarding the behavior of 
a magnetic field in a conducting medium that finds itself in a flat space time. 
Constructing for instance the field ${\vec B}_{1}(t,r,{\theta})$ corresponding 
to $S_{1}(t,r)$, one immediately sees that an observer at fixed $(r,{\theta})$ finds that
the magnitude of  ${\vec B}_{1}(t,r,{\theta})$ decays exponentially with a characteristic 
e-folding time given by ${\tau}_{1}={4{\sigma}R^{2}\over {\pi}c^{2}}$.
On the other hand expanding an arbitrary initial field configuration
${\vec B}(t_{0},r,{\theta})$ in terms of the eigenfunctions $({\psi}_{n}, n=1,2..)$, 
one can easily see the spatial diffusion of the initial distribution.
For a plasma characterized by an arbitrary ${\sigma}$, although the decay and diffusive 
nature of the initial ${\vec B}$ field remain intact, it is rather difficult to estimate
analytically the characteristic decay time as well as to find out whether the decay will 
be channeled into an exponential phase. 
It is sufficient, however, to stress that as long as we are in flat spacetime the decay 
process is controlled by the conducting properties of  the background medium and, of 
course, the length scale of the initial field distribution.

Let us now turn the discussion to the formulation of the  appropriate conditions to be 
imposed on solutions of eq.~\ref{20}.
Since as already indicated in the introduction section, a dipole field on a $Schwarzschild$ 
is modified considerably from its flat form and, as eq.~\ref{20} shows, relativistic effects
modify the local behavior of the relativistic Stoke's function ${\hat F}(t,r)$, one expects
modification of the boundary conditions as well.
As far as the behavior of ${\hat F}(t,r)$ at the star's center is concerned, by arguing 
in the same manner as in the flat space case,  a non singular  dipole field requires
${\hat F}(t,r)$ to satisfy identical conditions at the star's center as its flat counterpart, 
namely $lim{{\hat F}(t,r)\over r^{2}}$ should be finite as the center of the star is 
approached  (after all, the principle of equivalence holds).
Although this is the case at the star's center the corresponding boundary conditions at 
the star's surface are markedly different.
Recalling that the frame component of the   vector potential 
$A=A_{\mu}e^{\mu}=A_{\phi}e^{\phi}$ describing a magnetic dipole on a $Schwarzschild$
background is described by \cite{10}:
\begin{equation}
A_{\phi}={3{\mu} \sin{\theta}\over 4M^{2}}
\left[x \, \ln\left(1-{1\over x}\right)+{1\over 2x}+1\right]
\label{23a}
\end{equation}
where $x={r\over 2M(R)}$, $R~{\leq}r<{\infty}$, and ${\mu}$ is the dipole moment. 
From \ref{23a}, one then obtains the corresponding field 
${\vec B}=B^{r} \; \vec{e}_{r}+B^{\theta}\; \vec{e}_{\theta}$ where the physical
components $(B^{r}$, $B^{\theta})$ as  measured by the Killing observers are given by:
\begin{mathletters}
\label{23}
\begin{equation}
B^{r}={2{\mu} \cos{\theta}\over r^{3}}
      \left[3x^{3}\, \ln(1-x^{-1})+3x^{2}+{3\over 2}x\right]
\label{23b}
\end{equation}
\begin{equation}
B^{\theta}=-{{\mu} \sin{\theta}\over r^{3}}
\left[6x^{3}(1-x^{-1})^{1\over 2}\, \ln(1-x^{-1})+
6x^{2}{1-{1\over 2x}\over (1-x^{-1})^{1\over 2}}\right]
\label{23c}
\end{equation}
\end{mathletters}
A comparison of (3.17, ab)) with the corresponding eq.~\ref{16} and a $C^{0}$ matching between
the two along  the star's surface, requires that the exterior dipole magnetic moment ${\mu}$
should be identified with the generally slow time varying part of the function 
${\hat F(t,r)}$ \cite{25}. 
Moreover the gradient ${\hat F(t,r)}$ along the radial direction should obey: 
\begin{mathletters}
\label{24}
\begin{equation}
R \; \left. {\partial {\hat F(t,r)}\over {\partial r}}\right|_{R}=G(y)
{\hat F(t,R)}
\label{24a}
\end{equation}
with:
\begin{equation}
G(y)=y\; {{2y \ln(1-y^{-1})+{2y-1\over y-1}}\over y^{2} \ln(1-y^{-1})+
          y+{1\over 2}}
\label{24b}
\end{equation}
\end{mathletters}
and $y={R\over 2M(R)}$.

Thus the behavior  of the interior dipole field in the presence of curvature is described 
by ${{\hat F(t,r)}}$ satisfying the differential eq.~\ref{20}, subject to boundedness of 
${{\hat F}(t,r)\over r^{2}}$ as the star's center is approached, and additionally obeying  
\ref{24a} at its surface. 
Before we turn our discussion to the construction of solutions of eq.~\ref{20} 
 subject to the above discussed conditions, we would 
like to write  an explicit formula for the time evolution of ${\hat F(t,r)}$ under the 
assumption that geometry of the spacetime corresponds to a static spherically symmetric 
star, solution of Einstein's equations.
We may recall that in the derivation of eqs. (3.11) and (3.18,ab) we assumed an arbitrary
non-singular, static, spherically symmetric background geometry with the only constraint that
it joins smoothly to an exterior $Schwarzschild$ field across the surface of the star. 
Nowhere in the derivation we needed the explicit form of the $M(r)={Gm(r)\over c^{2}}$ nor 
the form of  $Z=Z(r)$. 
Hereafter we shall become more explicit and shall take the background interior geometry
to be a non singular solution of the coupled Einstein-perfect fluid system.
As is well known, and under the assumption that the background Maxwell field makes 
negligible contribution to the structure of the star \cite{27},  static spherically 
symmetric, perfect fluid solutions of Einstein's equations imply satisfaction of the 
following differential equations between the metric functions ${\Phi}(r)$, $M(r)$, the 
hydrostatic pressure $P(r)$, and mass density ${\rho(r)}$ (see for instance \cite{12,13}):
\begin{mathletters}
\label{25}
\begin{equation}
{d{\Phi}\over dr}={{M(r)+4{\pi}r^{3}P(r)}\over r^{2}\left(1-{2M(r)\over r}\right)}
\label{25a}
\end{equation}
\begin{equation}
{dM(r)\over dr}=4{\pi}r^{2}{\rho}
\label{25b}
\end{equation}
\begin{equation}
{dP(r)\over dr}=-({\rho}+P){M(r)+4{\pi}r^{3}P\over r^{2}
\left(1-{2M(r)\over r}\right)}
\label{25c}
\end{equation}
\end{mathletters}
Making use of those equations, and restoring the fundamental units,
we obtain from \ref{20} the following equation to be satisfied by the relativistic 
Stokes function:
\begin{equation}
{4{\pi}{\sigma}\over c^{2}}e^{-{\Phi(r)}}{\partial {F}\over {\partial t}}=
\left(1-{2Gm(r)\over c^{2}r}\right){\partial^{2}F \over {\partial r^{2}}}+
{1\over r^{2}}\left
[{{2Gm(r)\over c^{2}}+
{4{\pi}G\over c^{2}}r^{3}}\left({P(r)\over c^{2}}-{\rho(r)}\right)\right]
\, {\partial F\over {\partial r}}-
{2\over r^{2}}F
\label{26}
\end{equation}
where for typographical convenience we shall write here after $F(r,t)$ instead of 
${\hat F(r,t)}$. 
The above equation via \ref{16}, describes the evolution of 
any axisymmetric, dipole, poloidal
field ${\vec B}$ that finds itself interior to  a spherical perfect fluid star.
In the above form it includes all three relativistic factors influencing the field decay.
Since the distributions of $(m(r),~~ P(r),~~{\rho(r}))$ are related via Einstein's equations 
directly to the spacetime curvature, eq.~\ref{26} shows implicitly that 
the influence of 
spacetime curvature on the decay of the magnetic field is a real effect and cannot be 
removed via coordinate transformations.
In principle one could insert in \ref{26} the appropriate distributions of 
$m(r),~~ P(r),~~{\rho(r})$ resulting from integrating the Oppenheimer-Tolman-Volkov equation,
specify ${\sigma}={\sigma}(r,t)$, and construct the history of the ${\vec B}$-decay. 
We shall report elsewhere  our findings of this  rather laborious numerical 
integration \cite{28}.
For the purpose of the present paper we shall integrate \ref{26} for a rather simple 
system introduced and discussed in the following section.
The goal of this section is to show that the general relativistic eq.~\ref{26} (or its 
approximate forms corresponding to eq.~ (3.13b) under the assumption of a uniform
conductivity admits decay modes analogous to those of the flat space case with one 
important difference: 
The corresponding e-fold decaying times are  longer in the relativistic case. 
We interpret this amplification of the e-fold decay times as resulting from the non 
vanishing space time curvature.
Unfortunately however it is not easy to construct analytically 
the exact decaying modes 
of the full  relativistic system \ref{26} or its approximate versions
(3.13a), and thus  we shall
resort to numerical computations.
The emphasis in those computations is 
the probing of the dependence
of the corresponding e-folding times upon the value of the red shift factor or (and) 
upon the strength of the curvature of the spatial sections.

%%%%%%%%%%%%%%%%%%%%%%%%%%%%%%%%%%%%%%%%%%%%%%%%%%%%%%%%%%%%%%%%%%%%%%%%%%%%%%%%
\section{Magnetic field decay in a constant density star. Explicit results.}%%%%
%%%%%%%%%%%%%%%%%%%%%%%%%%%%%%%%%%%%%%%%%%%%%%%%%%%%%%%%%%%%%%%%%%%%%%%%%%%%%%%%

We shall consider in this section the decay of a magnetic field in a neutron star
of constant density. 
The assumption of a constant density star, although not a very reliable approximation of 
a real neutron star, offers the advantage that the Einsteins equations 
can be solved analytically (see for instance \cite{12,13}), and thus provides us with closed 
form expressions for the  coefficients of the induction equation, eq.~\ref{26}, and the 
boundary condition, eq.~\ref{24}.
In particular, in Box 23.2  of the ref.~\cite{13} the  distribution of the various 
hydrodynamical and geometrical variables are plotted as functions of the areal coordinate $r$. 
As we have already discussed, the exact decaying modes of the flat space-time induction 
equation for a uniform conductivity are explicitly known (given by eq.~\ref{22b}), while 
the corresponding decaying modes of the full curved space-time equation \ref{26} are 
presently unknown.
We shall therefore resort to numerical techniques in an attempt to get insights into 
behavior of the space of solutions of eq.~\ref{26}.

Viewed as an initial-boundary value problem,  (3.20) is a diffusive initial value problem 
for which the standard numerical technique is the Crank-Nicholson implicit integration 
scheme (see for example ref.[29] for a description).
We checked our numerical code by evolving the fundamental mode of the flat space-time 
case (ie take n=1 in eq.~\ref{22b}) and compare the numerical solution with the analytical 
one: we obtained an accuracy better than 1\% until times up to $10 \cdot {\tau}_1$, with 
${\tau}_{1}={4{\sigma}R^{2}\over {\pi}c^{2}} = \frac{\tau_{\rm Ohm}}{\pi^2}$  
the corresponding decay time of the $n=1$ flat fundamental mode.

Before we turn our discussion of the 
numerical results
it is helpful 
to view the time evolution of a  chosen initial distribution from a complementary 
point of view. 
On general grounds $F(t=0,r)$  as well as its time evolution can be (formally) expanded in 
a series of the following form:
\begin{equation}
F(t,x)={\Sigma}a_{n}e^{-{c^{2}{\lambda_{n}t}\over
4{\pi}{\sigma}R^{2}}}g_{n}(x)
\label{27}
\end{equation}

where the summation is extended over all eigenmodes $g_{n}(x)$ of the 
corresponding 
(singular) Sturm-Liouville eigenvalue problem arising from eq.~\ref{26} 
and the associated 
boundary-regularity conditions:

\begin{mathletters}
\label{28}
\begin{equation}
Lg_{n}+{\lambda}_{n}e^{-{\Phi}}g_{n}=0
\label{28a}
\end{equation},
\begin{equation}
 L=\left(1-{2Gm(x)\over c^{2}x}\right) {\partial^{2} \over {\partial x^{2}}} +
{1\over x^{2}}
\left[{{2Gm(x)\over c^{2}}+{4{\pi}G\over c^{2}}x^{3}}\left({P(x)\over c^{2}}-{\rho}\right)\right]
{\partial \over {\partial x}}-
{2\over x^{2}}
\label{28b}
\end{equation}
\end{mathletters}
where  in the present case the coefficients in $L$ are determined by the geometrical 
and hydrodynamical variables of the  constant density star solution, $x={r\over R}$ and
$R$ is the areal radius of the star. It follows now 
from (4.1) that if the eigenvalues are positive and well spaced,
then after $t>>{t_{Ohm}\over {\lambda_{1}}}$ where ${\lambda_{1}}$ 
is the lowest
eigenvalue of the above system, then the evolution of 
the distribution will channel into an exponentially decreasing 
phase with the dominant contribution in the sum (4.1) coming from 
the "first" term. Our subsequent described numerical computations exhibits such feature
and this property allows to construct numerically the lowest eigenvalue
of the above system~\cite{A}.

In the following numerical calculations we have taken the areal radius to be $R = 10$ km, 
a constant uniform conductivity ${\sigma} = 10^{25}$ s$^{-1}$ typical of neutron star 
values (which implies ${\tau}_{\rm Ohm} = 4.44 \; 10^9$ yrs), and we consider various 
neutron star masses characterized by different values of the dimensionless compactness ratio:
${\epsilon}={2GM\over Rc^2}$ = 0, 0.3, 0.4, 0.5, 0.6, 0.740, 0.810, 0.865, and 0.889.
The first one corresponds to a flat background  space time, the last four are those values 
used in the numerical plots of  ref. \cite{13}, while current
realistic neutron star models are characterized by ${\epsilon}$ in the range 
0.3 to 0.5 \cite{Prakash}. 
For each value of ${\epsilon}$, at first we  have solved numerically the full 
relativistic induction equation \ref{26}, by taking the initial $F(t=0,r)$ to be equal to 
the Stoke's function $S(t=0,r)$ of the corresponding first fundamental decay 
modes of \ref{22b} 
(ie take $n=1$ and $t=0$ in \ref{22b}). 
After performing a long time-integration of eq.~\ref{26} subject to the conditions 
cited earlier on, we find that the evolution of $F(t,x)$ channels into an 
exponentially decaying  mode
which  means, according to \ref{27}, that the evolution of the 
initial distribution eventually 
is described by the first non vanishing term in the series  expansion \ref{27}.
This
behavior of $F(t,x)$ allows us  to determine only the lowest eigenvalue
${\lambda}_{1}$ of (4.2) from our numerical outputs.
Besides the explicit determination of ${\lambda}_{1}$, our numerical 
treatment allows 
us to construct  the magnetic field as well. 
In Fig.~\ref{fig:1}, we plot as a function of coordinate time $t$, the magnetic
field as perceived by a Killing observer located at the star's pole for the various
values of the compactness ratio. 
Fig.(1) shows that
once curvature effects are incorporated and upon ignoring the initial transit time during 
which the field is in a superposition of various curved eigenmodes, the field follows 
an exponential decay law (as would have done in the absence of gravity) but now the 
corresponding e-folding time is longer than the corresponding flat spacetime case. 
Thus even though we have started with identically prepared systems 
their evolution is distinct,
a distinction traced in the influence of relativistic effects.
It should be stressed however that the content of
Fig.1 does not by itself provide us with a 
clear 
overall
picture of the field decay. It rather provides us 
with a characteristic physical decay time as 
perceived by a Killing observer situated at the surface of the star
and this decay time should not be extrapolated as being the physical decay time
 over the entire star ~\cite{31}.
In fact, each Killing observer located at some $r$, 
will
compute a physical e-fold decay time ${\tau(r)}$ given by: 
${\tau(r)}=Z(r)({\lambda}_{1})^{-1}={Z(r){\tau}_{Ohm}\over {\beta}{\pi^{2}}}$
and obviously this value changes across the star. (In this formula, we have 
parametrized ${\lambda}_{1}$
so that ${\beta}=1$ corresponds to flat spacetime).
Because of this spatial dependence of ${\tau}(r)$, in order 
to get a better insights into the dynamics of 
the decay, in Fig.2, we have plotted  
${\lambda_1}={\beta}{{\pi}^{2}\over \tau_{\rm Ohm}}$,
as a function of the compactness ratio ${\epsilon}$. In the same figure,
for comparison purposes, we have plotted the value of the red shift
factor at the stars center ($Z_{o}=e^{\Phi(o)}$) and surface 
($Z_{s}=e^{\Phi(s)}$) respectively.
Thus it follows from Fig.(2),
that the relativistic corrections to the lowest eigenvalue ${\lambda}_{1}$,
are
bounded from above by $Z_{s}$ while from bellow (almost) by $Z_{0}$.
It is more instructive however and complements the content of Fig.(2), a
plot showing the physical decay times as measured by Killing observers 
located at the center and at the surface of the star 
respectively. Fig.(3) stands for such plot, and its content shows that 
the physical decay time can vary
considerably across the star. In fact there are regions
around the star's center, where  the physical decay 
time is shorter than the corresponding flat
space case and this effect is more pronounced as 
the compactness ratio increases.
In contrast to what occurs in the vicinity of the star's center, in the crust
region the physical decay
is always larger than the corresponding flat case.
Because of this behavior, largely due to gravitational time dilation
effect, we assign an  overall physical decay  time, by 
averaging the physical
e-fold decay at the center 
and the surface of the star respectively.
This amounts to
assigning an  overral a red shift
factor $Z$ for the entire star
equal roughly to its value at the middle of the star. 
With this type of averaging,
Figs.(2,3) shows
that for  small
values of the compactness ratio
the overall physical decay time
is almost identical to the flat space time case.
However, as the compactness ratio increases, the relativistic effects become
more apparent.
For the case of neutrons stars with range in the realistic domain
ie ${\epsilon}$ in the range $(0.3~,~0.5)$,
and via the averaging procedure outlined above, the overall 
physical e-fold decay time
is $(1.2-1.3)$ larger than the corresponding 
flat case. 
Although the content of Fig.~\ref{fig:1}, Fig.~\ref{fig:2} 
and Fig.~\ref{fig:3}
show the impact of 
relativistic effects upon the
field decay, by themselves they do not offer a clear insight as which (if any) 
of the two factors, 
ie red shift or spatial curvature, are responsible for the 
dominant contribution in the 
field decay.
In order to access their relative importance  we solve 
numerically eq.~\ref{21a} 
in two extreme cases~\cite{B} and show the numerical outputs in Fig.(4).
First  eq.~\ref{21a} is solved
under the assumption $r(l)=l$ and 
in this approximation the relativistic effects on the decay are solely 
due to the 
red shift factor $Z(l)$. In fig.(4) the numerical outputs are
indicated by the label: "curved time".
In the opposite extreme, we adopt  $Z(l)=1$ in eq.~\ref{21a}
and take $r(l)$ as given by the metric 
corresponding to a constant density star. Thus
in this approximation,  the only relativistic effect influencing the decay
is due to the 
spatial curvature. The resulting numerical outputs
in Fig.(4) are 
marked by the label "curved space". 
It follows then clearly from the content 
of Fig.(4) that for a constant density neutron star,
the dominant effect in  the field decay is due
 to the red-shift factor $Z$, since the corresponding
eigenvalues  indicated by "curved-time" graph are much
closer to the corresponding exact eigenvalue
indicated by "curved space-time" in Fig.(4). 
Moreover, the dominance of the red-shift factor holds through
for all values of the compactness ratio 
${\epsilon}$, and increases as ${\epsilon}$ increases.

From the analysis presented so far it is clear that the 
 more compact the star is, the longer is the 
e-folding time.
As a consequence  one expects that models of pulsars with soft equation of state to maintain 
a strong magnetic field for longer period of time than the corresponding models with a stiff 
equation of state.
In turn such slow $B$ field decay implies additional source of heating ie Joule heating and 
such additional heating, may explain the relatively high temperature observed in old neutron
stars.
However based on the present analysis, it is rather premature to draw definite conclusions.
For instance cooling effects leading to the temporal variation of the conductivity as well
as the the detailed structure of the star and its rotation has to be taken into
account.
Such study is currently under way and we expect to report in a  future communication.

%%%%%%%%%%%%%%%%%%%%%%%%%%%%%%%%%%%%%%%%%%%%%%%%%%%%%%%%%%%%%%%%%%%%%%%%%%%%%%%%
\section{Conclusions}%%%%%%%%%%%%%%%%%%%%%%%%%%%%%%%%%%%%%%%%%%%%%%%%%%%%%%%%%%%
%%%%%%%%%%%%%%%%%%%%%%%%%%%%%%%%%%%%%%%%%%%%%%%%%%%%%%%%%%%%%%%%%%%%%%%%%%%%%%%%

The  behavior of the surface magnetic fields of neutron stars is a complicated and 
controversial issue.
Many processes are believed to influence its magnitude and its subsequent evolution. 
Trapping for instance, of the field in the superconducting core is one possibility. 
The expulsion of the field out of this region is a delicate matter involving many different
branches of physics \cite{32}. 
Another possibility that in principle influences enormously the magnetic properties of neutron 
stars is related to the accretion processes immediately after the core collapse \cite{33}.
Accretion and particularly hypercritical accretion, can submerge the field of the new born 
neutron star beneath a layer of accreting matter thus in principle producing a delayed 
switched on mechanism for the pulsar activity \cite{34}. 
Furthermore according to recent work  the neutron star may never turn on as a pulsar 
\cite{35} if the accretion is hyperctical. 
Besides the above mechanisms influencing the evolution of neutron star's magnetic fields, 
many more have been introduced and discussed at length in the current literature.
In this work, we have present a limited framework
taking into account the effects of space time curvature
 on the field decay.
For the simple neutron star models with a corresponding compactness
ratio in the range $(0.3~,~0.5)$, considered in the present work,
we have seen an overall increase in the decay time, $(1.2$ to $1.3)$
times larger than the flat spacetime value.
Although the present work is preliminary and to assess the new effect more work is 
needed \cite{28}, it  point towards to the direction 
that in a strongly gravitating system, 
effects due to space time curvature should not be neglected.

\acknowledgments%%%%%%%%%%%%%%%%%%%%%%%%%%%%%%%%%%%%%%%%%%%%%%%%%%%%%%%%%%%%%

This work was supported by a binational grant
DFG (grant \#444 - MEX - 1131410) -
Co\-na\-cyt (grant \#E130.443),
Co\-na\-cyt (grant \#2127P - E9507),
UNAM - DGAPA (grant \#IN105495)
and Co\-ordi\-na\-ci\'on Cien\-t\'{\i}\-fi\-ca - UMSNH. 
D.P. and T.Z. are thankful to the
Astro\-phy\-si\-ka\-li\-sches Inst\-it\-ut Potsdam for its kind hospitality and
U.G. to the Ins\-ti\-tu\-to de As\-tro\-no\-m\'{\i}a of UNAM.

\appendix

%%%%%%%%%%%%%%%%%%%%%%%%%%%%%%%%%%%%%%%%%%%%%%%%%%%%%%%%%%%%%%%%%%%%%%%%%%%%%%%%
\section{$(3+1)$ Form of Maxwell's Eqs. on Static spacetimes}%%%%%%%%%%%%%%%%%%%%%%%%%%%%%%%%%%%%%%%%%%%%%%%%%%%%%%%%%%%%%%
%%%%%%%%%%%%%%%%%%%%%%%%%%%%%%%%%%%%%%%%%%%%%%%%%%%%%%%%%%%%%%%%%%%%%%%%%%%%%%%%

%\appendix%%%%%%%%%%%%%%%%%%%%%%%%%%%%%%%%%%%%%%%%%%%%%%%%%%%%%%%%%%%%%

In this Appendix we shall sketch a derivation of eqs.(2.8) starting from the covariant 
form of Maxwell's
eqs.~\ref{2}. 
The derivation makes use of the existence of the hypersurface orthogonal timelike Killing
field and although all the following  computations can be done in a covariant fashion 
\cite{18,20}
for 
brevity  we work explicitly in the coordinate gauge of \ref{5}.  
We shall also present formulas required for the derivation 
of equations in section II.

Starting from the temporal component of the inhomogeneous Maxwell eq.~\ref{2a}, combined with 
the line element \ref{5} and 
taking into account the fact that $E^{j}=-F^{jo}e^{{\Phi}}$
one immediately obtains:
\begin{equation}
D_{i}E^{i}=-{4{\pi}\over c}J^{o}e^{\Phi}=4{\pi}{\rho}\label{A1}
\end{equation}
where we have defined the charge density ${\rho}$ measured by a Killing observer by: 
$c{\rho}=-J^{\mu}U_{\mu}$ and we denote by $D$  the covariant derivative 
operator associated with the Riemannian metric of the $t=const$ spaces.
Due to the fact that 
the Maxwell tensor $F_{\alpha\beta}$ admits the following easy verifiable decomposition: 
$F_{{\alpha\beta}}=U_{\alpha}E_{\beta}-U_{\beta}E_{\alpha}+
{\epsilon}_{\alpha\beta\gamma\delta}U^{\gamma}B^{\delta}$
one gets the following expression for its spatial part $F_{ij}$:
$F_{ij}={\epsilon}_{oijl}B^{l}U^{o}={\epsilon}_{ijl}B^{l}$.
Passing now to the spatial components of \ref{2a} one gets 
\begin{equation}
{\partial E^{i}\over {\partial x^{o}}}-{\epsilon}^{ijl}D_{j}(ZB_{l})=
-{4{\pi}\over c}ZJ^{i}
\label{A2}
\end{equation}
where we have introduced the red shift factor $Z$ via
$Z=(-{\xi}^{a}{\xi}_{a})^{1\over 2}=e^{\phi}$ instead of $e^{\Phi}$.
On the other hand the second pair of Maxwell's equations \ref{2b} can be written 
equivalently as:
\begin{equation}
{\partial F_{\mu\nu}\over {\partial x^{\lambda}}}+
{\partial F_{\nu\lambda}\over {\partial x^{\mu}}}+
{\partial F_{\lambda\mu}\over {\partial x^{\nu}}}=0
\end{equation}
Taking now all the indices to be spatial, and eliminating $F_{ij}$  one gets:
$D_{i}B^{i}=0$
The other information encoded in the second pair of Maxwell eqs. can be revealed by 
considering  the following arrangement of the spacetime indices: 
$({\mu}, {\nu}, {\lambda}=m, n, x^{o})$. 
For such arrangement one obtains:
\begin{equation}
{\partial ({\epsilon}_{mnl}B^{l})\over {\partial {x^{o}}}}+
{\partial (-U_{o}E_{n})\over {\partial x^{m}}}+
{\partial (U_{o}E_{m})\over {\partial x^{n}}}=0
\end{equation}
from which one easily obtains:
\begin{equation}
{\partial B^{l}\over {\partial {x^{o}}}}+{\epsilon}^{lmn}D_{m}(ZE_{n})=0
\label{A5}
\end{equation}
The current conservation eq. ${\nabla}_{\mu}J^{\mu}=0$ after a trivial rearrangement yields:
\begin{equation}
U^{\mu}{\partial c{\rho}\over {\partial x^{\mu}}}+D_{i}J^{i}+J^{i}{\partial
logZ\over {\partial x^{i}}}=0
\label{A9}
\end{equation}

To pass into the equivalent set (\ref{9}-abcd) and eq.~\ref{10} involving physical 
orthonormal components, we project all tensors involved onto the  natural set of orthonormal 
vectors $(e_{i})$ and one forms $(e^{i}),~~~(i=1,2,3)$ respectively, associated with the 
line element \ref{8}.
Thus for instance the electric field $E$ can be written as:
$E=E^{i}{\partial \over {\partial x^{i}}}=E^{{\hat i}}e_{i}$
where $E^{{\hat i}}=h_{i}E^{i}$ expresses the relationship between coordinate and frame 
components of the vector field $E$ and it is understood that no summation is involved over 
the repeated indices.
With the help of the orthonormal components one for instance  may rewrite eq.~\ref{A1} in 
terms of orthonormal components.
Writing ${\gamma}^{1\over 2}=h_{1}h_{2}h_{3}$ and eliminating the coordinate components in 
terms of the frame components of $E$, equation  ~\ref{A1} takes the following form:
\begin{equation}
{1\over h_{1}h_{2}h_{3}}
[{{\partial \over {\partial x^{1}}}(h_{2}h_{3}E^{{\hat 1}})
+{\partial \over {\partial x^{2}}}(h_{1}h_{3}E^{{\hat 2}})
+{\partial \over {\partial x^{3}}}(h_{1}h_{2}E^{{\hat 3}})}]=
{\nabla}{\cdot}{\vec E}={4{\pi}}{\rho}
\label{A10}
\end{equation}
where ${\nabla}{\cdot}$ stands for the familiar divergence operator expressed in arbitrary 
orthogonal curvilinear coordinates 
defined by the line element \ref{8}.
Similarly $D_{i}B^{i}=0$
can be written as ${\nabla}{\cdot}{\vec B}=0$.
As far as the other set of Maxwell's eqs. are concerned, one can 
proceed in a similar manner.
For instance starting from A2, one first multiplies the corresponding equation 
by the scale factor $h_{i}$ thus leading into:
\begin{equation}
{\partial (E^{{\hat i}}) \over {\partial x^{o}}}-
{{\epsilon}^{{\hat i}{\hat j}{\hat k}}h_{i}\over
h_{1}h_{2}h_{3}}{\partial (h_{k}B_{{\hat k}}Z)\over {\partial x^{j}}}=-{4{\pi}\over c}
J^{{\hat i}}Z
\label{A11}
\end{equation}

Recalling that the orthonormal components of the ${\it curl}$ operator of an arbitrary 
three dimensional differentiable vector field $A$ are given by \cite{36}:
\begin{equation}
({\nabla}{\times}A)^{\hat i}={{\epsilon}^{{\hat i}{\hat j}{\hat k}}h_{i}\over
h_{1}h_{2}h_{3}}{\partial (h_{k}A_{\hat k})\over {\partial x^{j}}}
\label{A12}
\end{equation}
one is lead immediately into eq.~\ref{9b} used in the text.
Note also that the action of the gradient operator ${\nabla}$ acting on 
scalars is defined via:
\begin{equation}
{\nabla}f=
 {1\over h_{1}}{\partial f \over {\partial x^{1}}}+
 {1\over h_{2}}{\partial f \over {\partial x^{2}}}+
 {1\over h_{3}}{\partial f \over {\partial x^{3}}}
\end{equation}
Also in deriving eqs. (2.8-abc) of the main text we have used the following properties of 
the unit basis vectors $(e_{i})$:
$e_{\hat i}  {\cdot}  e_{\hat j} = {\delta}_{{\hat i}{\hat j}}$,
$e_{\hat i} {\times } e_{\hat j} = {\epsilon}_{{\hat i}{\hat j}{\hat k}}e_{\hat k}$
%=e_{{\hat i}{\hat j}{\hat k}}e_{{\hat k}}$
and the normalization ${\epsilon}_{{\hat r}{{\hat \theta}}{{\hat \phi}}}=1$.
We may also indicate  that for typographical convenience the caret-symbol over frame 
components of the various tensors has been dropped.
In particularly all vector, tensor components appearing anywhere in the main text after 
eq.~\ref{8}, are frame components.

%%%%%%%%%%%%%%%%%%%%%%%%%%%%%%%%%%%%%%%%%%%%%%%%%%%%%%%%%%%%%%%%%

\begin{figure}%%%%%%%%%%%%%%%%%%%%%%%%%%%%%%%%%%%%%%%%%%%%%%%%%%%%%%%%%%%%%%%%%%%
\caption{Figure 1}
Dipolar field decay for a uniform density star in curved and flat space-time.
The horizontal axis represents (coordinate) time in units of the flat space-time
ohmic decay time $\tau_{ohm} \equiv 4 \pi \sigma R^2/c^2$ and the
vertical axis shows the value of
$B/B_0 = B(t,r=R, \theta =0)/B(t=0,r=R, \theta =0)$.
All models have the same areal radius ($R = 10$ km) and a constant
uniform conductivity ($\sigma = 10^{25}$ s$^{-1}$) typical of
neutron star values (which implies $\tau_{ohm} = 4.44 \; 10^9$ yrs).
The values of the compactness ratio
$2GM/Rc^2$ is indicated on each plot.
The initial field profile is taken as the $n =1$ eigenmode for the flat
space-time, Eq.\protect\ref{22b}, in all cases.
The graphs show quite clearly the exponential decay in flat space-time,
with $\tau = \tau_{ohm}/\pi^2$ within numerical accuracy, while, as expected, 
in curved space-time the decay initially deviates from an exponential law
but rapidly converges toward the corresponding fundamental mode.
\label{fig:1}
\end{figure}%%%%%%%%%%%%%%%%%%%%%%%%%%%%%%%%%%%%%%%%%%%%%%%%%%%%%%%%%%%%%%%%%%%%%

\begin{figure}%%%%%%%%%%%%%%%%%%%%%%%%%%%%%%%%%%%%%%%%%%%%%%%%%%%%%%%%%%%%%%%%%%%
\caption{Figure 2} The horizontal axis stands for the dimensionless
comactenes ratio $2GM/Rc^2$, while the vertical axis corresponds to
the values of ${\beta}$. The graphs marked by $e^{{\Phi}_{o}}~,~e^{{\Phi}_{s}}$
stand for the value of the red shift factor at the center and surface of the
star respectively. 
\label{fig:2}
\end{figure}%%%%%%%%%%%%%%%%%%%%%%%%%%%%%%%%%%%%%%%%%%%%%%%%%%%%%%%%%%%%%%%%%%%%%

\begin{figure}%%%%%%%%%%%%%%%%%%%%%%%%%%%%%%%%%%%%%%%%%%%%%%%%%%%%%%%%%%%%%%%%%%%
\caption{Figure 3}The horizontal axis stands for the dimensionless
compactenes ratio $2GM/Rc^2$, while the vertical axis corresponds to
the ratio of the physical decay time ${\tau}_{ph}$ over the corresponding
flat value ${\tau}_{fl}$. The graphs marked as "surface",
"center" respectively, represents
${{\tau}(R)_{ph}\over {\tau}_{fl}}={Z_{s}\over {\beta}}$,
${{\tau}(0)_{ph}\over {\tau}_{fl}}={Z_{o}\over {\beta}}$
while the corresponding horizontal line through $(0~,~1)$ stands for 
the flat case. 
\label{fig:3}
\end{figure}%%%%%%%%%%%%%%%%%%%%%%%%%%%%%%%%%%%%%%%%%%%%%%%%%%%%%%%%%%%%%%%%%%%%%

\begin{figure}%%%%%%%%%%%%%%%%%%%%%%%%%%%%%%%%%%%%%%%%%%%%%%%%%%%%%%%%%%%%%%%%%%%
\caption{Figure 4}The horizontal axis stands for the dimensionless
comactenes ratio $2GM/Rc^2$, while the vertical axis corresponds to
the values of $\beta$. The graphs marked as "curved space","curved time"
provides the eigenvalues corresponding the the case
where $Z=1$  and $r(l)=l$ respectively, as explained in the text,
while the graph marked as "curved space time" corresponds to the
exact equation.
\label{fig:4}
\end{figure}%%%%%%%%%%%%%%%%%%%%%%%%%%%%%%%%%%%%%%%%%%%%%%%%%%%%%%%%%%%%%%%%%%%%%

%%%%%%%%%%%%%%%%%%%%%%%%%%%%%%%%%%%%%%%%%%%%%%%%%%%%%%%%%%%%%%%%%%%%%%%%%%%%%%%%

\begin{references}
\bibitem{11} There are some sensitive issues involved in the use of the terms
 "influence of general 
relativistic effects" and  "$B$ field decay". It would be perhaps more
correctly to write that we are interested to study
how certain solutions of Maxwell's equations
expressed
relative to
global Lorentz frames and describing
a particular physical situation would appeared
once they the same physical problem
considered in the presence of non vanishing space time 
curvature. 
 As far as the term  "influence of general 
relativistic effects" is concerned here rather at the start 
we are interested in considering the effects of spacetime curvature. 
Maxwell's eqs. on flat space time can be written in an arbitrary 
curvilinear system and the issue  how a B-decaying field relative 
to inertial frames would be perceived by 
in a non inertial observers, naturally arises.
The present paper is not intended to analyze such issue. 
Toward to the end of the paper we shall make a few relevant comments 
pertinent to that issue.
\bibitem{14} A note on our conventions: Maxwell's eqs.(apart from the insertion of 
the c factor in eq.~\ref{2a}), signature of metric, curvature etc, are the same as that 
of ref. \cite{12}. 
In addition in the present work all Greek indices are assumed to be four dimensional, 
while Latin indices are spatials.
The definition and normalization of the four and three dimensional  totally antisymmetric
Levi-Civita tensor density is the same as that of ref. \cite{12}.
\bibitem{16} We shall ignore in the present work possible 
additional effects of high B-field
leading into a conductivity tensor. For simplicity we shall work with a 
scalar conductivity, 
leaving the complete treatment to future work.
\bibitem{19} The MHD approximation employed in the present paper is well suited for 
astrophysical plasmas and particularly for neutron star matter where typically 
${\sigma}{\geq}10^{25}sec^{-1}$. 
To see that let us suppose  that $({\vec E}, {\vec B})$ is a solution of \ref{9}abc 
subject to condition ${\vec J}={\sigma}{\vec E}$.
If by $R$ we denote the typical length scale of the system, then eq.~\ref{9c} implies 
that in order of magnitude, ${|E|\over |B|}={R\over cT}$, where $T$ is the dynamical
evolution time scale for the $E, B$ fields and in arriving at that  estimate we have 
taken $Z$ to be of order unity. 
On the other hand 
%
${\partial E \over {\partial x^{o}}}{1\over {\nabla}{\times }B}= ({R\over cT})^{2}$ 
%
and thus taking into account Ohm's law in eq.~\ref{9b} one obtains 
%
$1= {{4{\pi}{\sigma}R^{2}}\over c^{2}}{1\over T} +({R\over cT})^{2}$.
%
It follows then from this estimate, if the evolution time scale $T$ 
is of the order of the Ohmic time, ie  $T={4{\pi}{\sigma}R^{2}\over c^{2}}$
then for the typical neutron star parameters ${R\over cT}<<1$ and for such situations
the displacement current can be neglected from the right hand side of \ref{9b} implying 
that the MHD approximation is well justified.
Since essentially in the  MHD approximation one neglects the electromagnetic radiation, 
that point will allows us further on to join the interior solutions with static exterior 
dipole field.
\bibitem{21} Actually even for a time dependented ${\sigma}$, the toroidal and 
poloidal component evolve independently of each other. 
In the present work we shall use in section III a uniform conductivity, 
however as we go along we shall point out implications on the evolution of the magnetic 
field components due to a conductivity characterized by an arbitrary spacetime dependence.
\bibitem{22} In the case of an arbitrary l-pole field one recovers an identical equation 
as the above. 
The sole exception is that in the factor of 2, in last term of the right hand-side is 
replaced by $l(l+1)$ respectively.
\bibitem{25} Strictly speaking the interior field for both flat and curved space case,
should be joined with a radiating into empty space solution of Maxwell's eqs. 
However due to the MHD approximations and due to the long times evolved in the decay of the 
interior field typically and to a good approximation  one considers the exterior dipole field
to consist of a sequence of quasi static dipole solutions. 
Thus essentially we take the exterior dipole magnetic moment to be a slow varying function of
time and that approximation allows us to perform the $C^{o}$ mashing of the ${\vec B}$ across 
the star's surface.
\bibitem{A} It should be stressed  however that knowledge of all 
eigenvalues  yields information  regarding the
behavior of the magnetic field before the exponentially decreasing phase is reached.
Accordingly construction of the other eigenvalues is a worthwhile project.
\bibitem{31}It should be stressed here that even though
we are drawing physical conclusions based on the magnetic 
field measured by Killing observers,  the influence
of the curvature can be seeing and described in a coordinate 
and observer free
manner. One for instance 
may consider  the field invariant
$F^{\alpha\beta}F_{\alpha\beta}=2[({\vec B})^{2}-({\vec E})^{2}]$
and examine its properties on a flat and curved spacetime.
Within the MHD approximation, the ${\vec E}$ field can be computed
via Ohm's law and thus the right handside of this field invariant
is expressible in terms of the corresponding Stokes functions.
In fact for our case since the ${\vec E}$ and current ${\vec J}$,
are purely toroidal, they vanish
 on the poles of the star, hence
$[F^{\alpha\beta}F_{\alpha\beta}(t,R,{\theta}=0)]^{1\over 2}
{\simeq}|{{\vec B}}(t,R,{\theta}=0)|$.
Thus Fig.(1) also supplies  information regarding
the behavior of this field invariant for the curved-flat space
respectively.
Accordingly the "flat value" of 
$F^{\alpha\beta}F_{\alpha\beta}(t,R,{\theta}=0)$ decays more
rapidly than its curved counterpart.
We have chosen to indicate the effect in terms of the ${\vec B}$
since the latter is  directly related to terrestrial observations.
\bibitem{B} In our numerical integration of (3.13a) for the cases
$Z=1$ and $r(l)=l$ respectively, we have used as boundary condition
the expression (3.18). Although this is not entirely correct, numerically
we have found that the content of Fig.(4) is rather insensitive to  
small changes in the surface boundary condition.
\end{references}
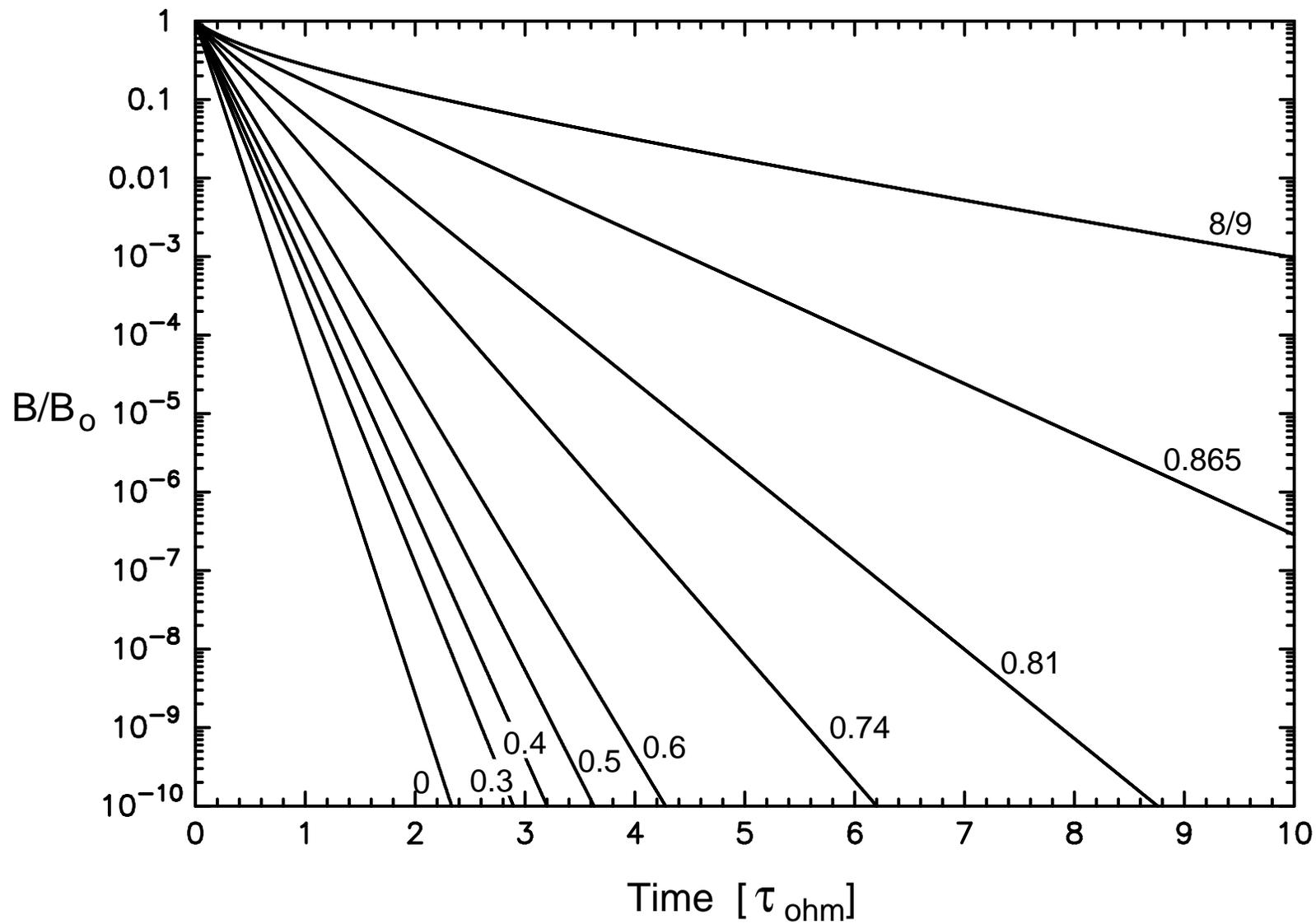
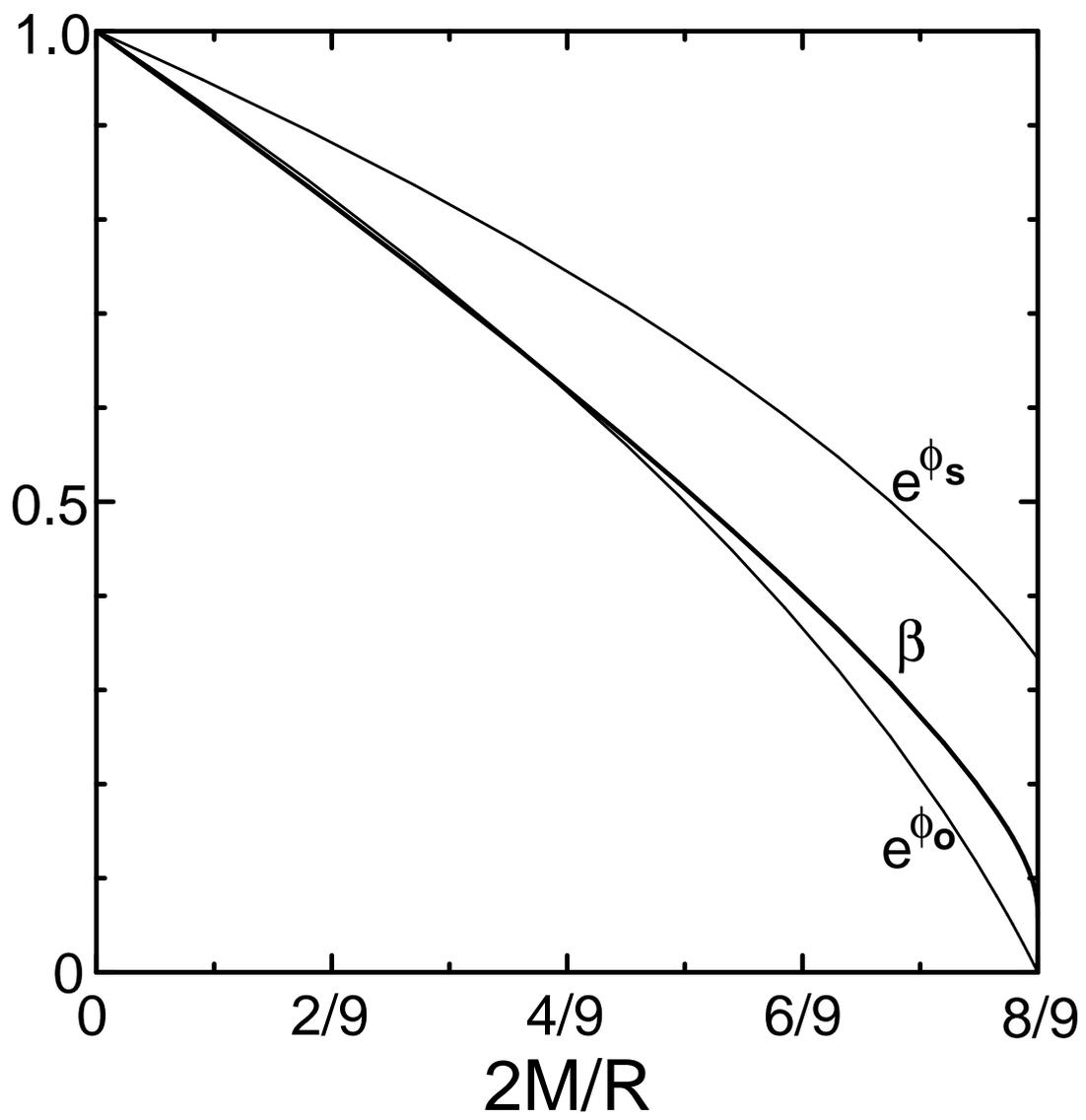
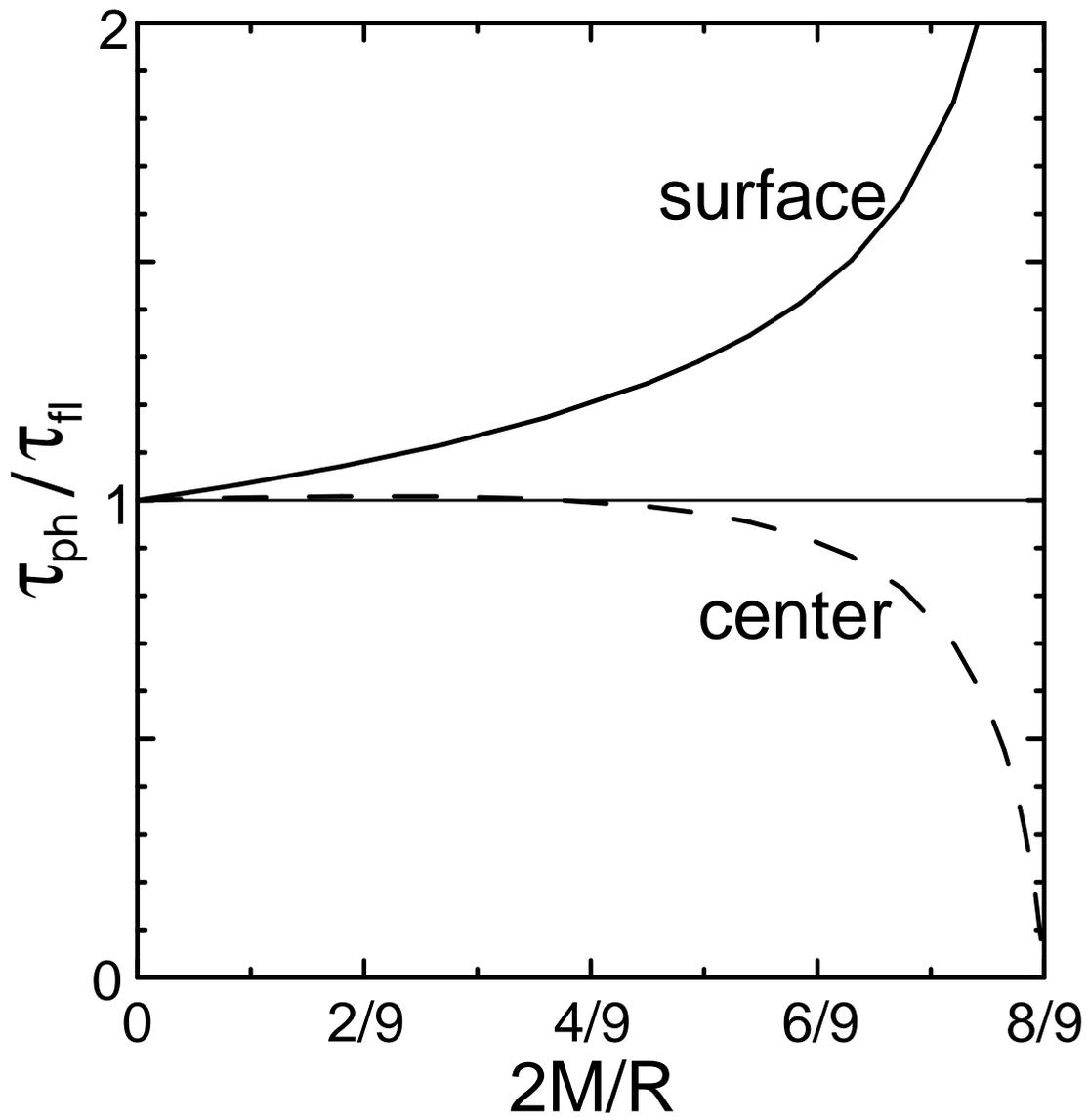
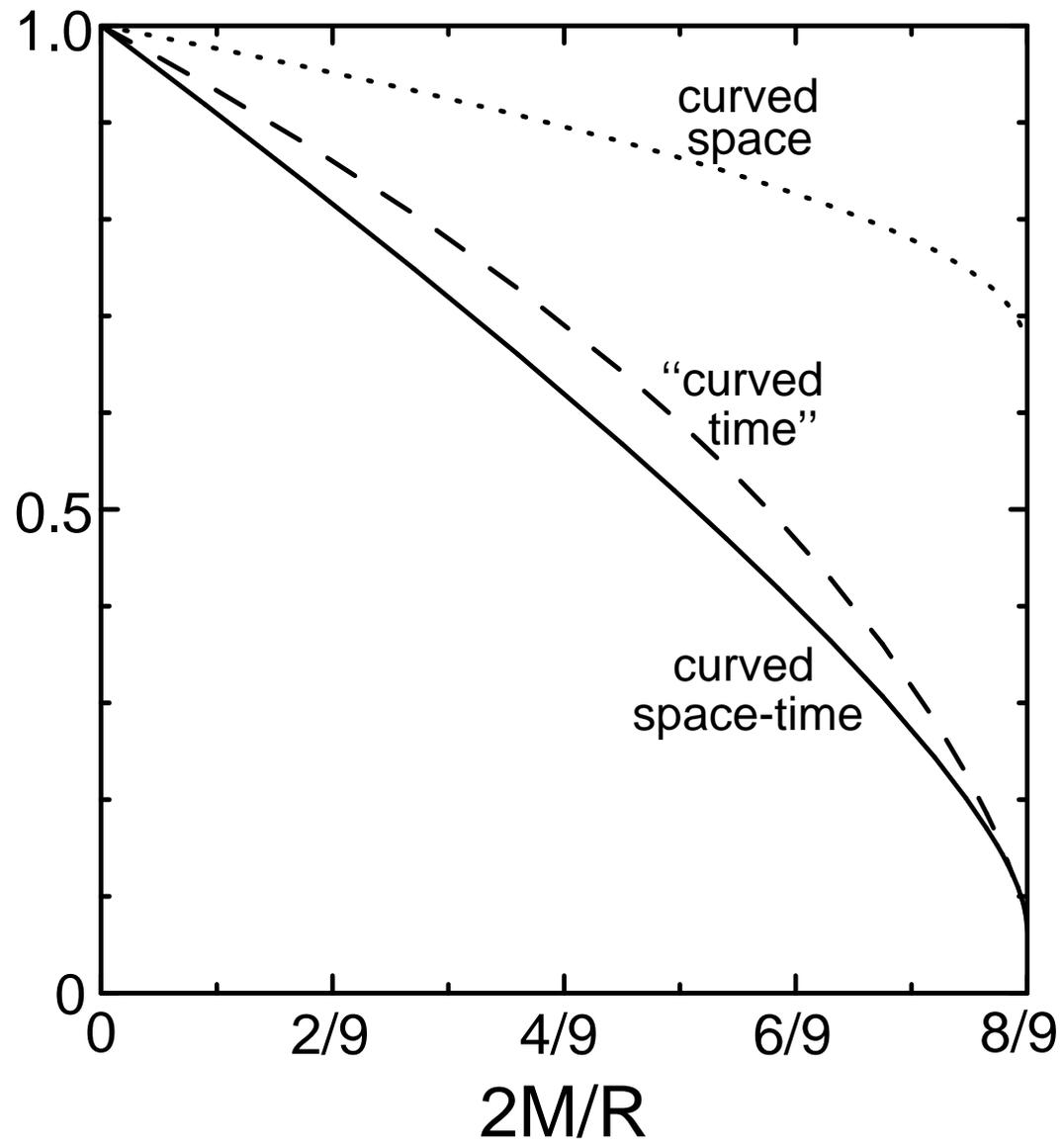
\end{document}